\documentclass[12pt,preprint]{aastex}

\usepackage{longtable}

\newcommand{\xmm}{{\it XMM-Newton}}
\newcommand{\wse}{{\it WISE }}

\newcommand{\swf}{{\it SWIFT}}

\newcommand{\chn}{{\it Chandra}}
\newcommand{\fer}{{\it FERMI}}
\newcommand{\inte}{{\it INTEGRAL}}

\shorttitle{New $\gamma$-Ray Blazar Candidates}
\shortauthors{Cowperthwaite et al. 2013}

\begin{document}


\title{Identification of New Gamma-Ray Blazar Candidates With Multifrequency Archival Observations}


\author{Philip S. Cowperthwaite\altaffilmark{1,2}, F. Massaro\altaffilmark{3}, R. D'Abrusco\altaffilmark{4}, A. Paggi\altaffilmark{4}, G. Tosti\altaffilmark{5,6}, \and Howard A. Smith\altaffilmark{4}} 


\altaffiltext{1}{Department of Astronomy, University of Maryland, College Park, MD 20742-2421, USA; pcowpert@umd.edu}
\altaffiltext{2}{Joint Space-Science Institute (JSI), College Park, MD 20742-2421, USA}
\altaffiltext{3}{SLAC National Laboratory and Kavli Institute for Particle Astrophysics and Cosmology, 2575 Sand Hill Road, Menlo Park, CA 94025, USA}
\altaffiltext{4}{Harvard--Smithsonian Center for Astrophysics, 60 Garden Street, Cambridge, MA 02138, USA}
\altaffiltext{5}{Dipartimento di Fisica, Universit\`a degli Studi di Perugia, 06123 Perugia, Italy}
\altaffiltext{6}{Istituto Nazionale di Fisica Nucleare, Sezione di Perugia, 06123 Perugia, Italy}


\begin{abstract}
Blazars are a highly-variable, radio-loud subclass of active galactic nuclei (AGN). In order to better understand such objects we must be able to easily identify candidate blazars from the growing population of unidentified sources. Working towards this goal we attempt to identify new $\gamma$-ray blazar candidates from a sample of 102 previously unidentified sources. These sources are selected from Astronomer's Telegrams and the literature on the basis of non-periodic variability and multi-wavelength behavior. We then attempt to associate these objects to an IR counterpart in the \wse\ all-sky survey. We are able to identify sixteen candidate sources whose IR colors are consistent with those of the blazar population. Of those sixteen, thirteen sources have IR colors indicative of being $\gamma$-ray emitting blazar candidates. These sources all possess archival multi-wavelength observations that support their blazar-like nature.
\end{abstract}


\keywords{galaxies: active --- galaxies: BL Lacertae objects ---  radiation mechanisms: non-thermal}
\clearpage

\section{Introduction}
\label{sec:intro}
Blazars are one of the most enigmatic classes of active galactic nuclei (AGN). The picture put forth by Blandford \& Rees (1978) suggested that blazars are the rare occurrence in which a relativistic jet, fueled by the accretion of material onto the central engine of the AGN, is beamed directly along the observer's line of sight. Observationally, blazars are generally characterized by rapid and strong variability, high and variable polarization, and a host of relativistic effects including superluminal motion and frequencies enhanced by the Doppler effect \citep[][]{urry95}. They also show peculiar infrared (IR) colors \citep[][]{dabrusco12, massaro11b}. 

We can further divide blazars into two distinct subclasses: BL Lac objects and flat spectrum radio quasars. It is convenient to utilize the nomenclature of the Multi-wavelength Blazar catalogue \citep[ROMA-BZ-CAT,][]{massaro09,massaro10,massaro11a}. We label the BL Lac objects as BZB objects and the flat-spectrum radio quasars as BZQ objects. BZB objects are characterized by featureless optical spectra, with weak or no emission lines \citep[e.g.,][]{stickel91,stoke91}. They also show a broadly peaked spectral energy distribution (SED) which appear to have two primary components: the first peaking in the IR to X-ray range and the second peaking in the $\gamma$-rays up to TeV energies \citep[e.g.,][]{padovani95}. BZQ objects are characterized by broad emission lines typical of the quasar optical spectra, with the high energy SED component peaking in the MeV--GeV range. Blazars which have insufficient data available to make this distinction are named ``blazars of uncertain type (BZU)" \citep[][]{massaro11a}.

Blazars feature high luminosity continua dominated by non-thermal emission. This emission extends from the radio to TeV $\gamma$-rays. As a matter of fact, blazars are the largest known population of $\gamma$-ray sources \citep[e.g.,][]{hartman99,abdo10a}. As a result, they are the dominant contributor to the extragalactic $\gamma$-ray background \citep[e.g.,][]{mukherjee97,abdo10b}. Additionally, blazars are  commonly detected in IR sky surveys. This was noted early on by the InfraRed Astronomical Satellite \citep[IRAS, e.g.,][]{impey88} and Wilkinson Microwave Anisotropy Probe \citep[WMAP, e.g.,][]{giommi07,giommi09}. Recent work with the Planck \citep{giommi12a} and HERSCHEL \citep{gonzales10} observatories confirms this result.

Making use  of the preliminary data release of the Wide-field Infrared Survey Explorer (\wse) 
\citep[see][for more details]{wright10}, 
Massaro et al. (2011a) found that $\gamma$-ray emitting blazars
occupy a distinct region of the 3-dimensional IR color-color space. This region is well separated from other extragalactic sources, particularly those dominated by thermal emission.
\citep[see also][]{dabrusco12}.

Based on this discovery, a new association procedure was developed \citep{dabrusco12,massaro12b} in order to search for $\gamma$-ray blazar candidates. This allowed one to search for $\gamma$-ray blazar candidates among the unidentified $\gamma$-ray sources present in the second \fer\ LAT catalog \citep[2FGL; ][]{nolan12}. This association procedure was also successfully applied to the 4$^{th}$ {\it INTEGRAL} catalog \citep{massaro12c}.

There have been several recent improvements to the association procedure, including a more conservative approach based on the \wse\ full archive
\footnote{http://wise2.ipac.caltech.edu/docs/release/allsky/} available since March 2012 \citep[see also][]{cutri12, paper6, massaro13}. In the following, we will make use of the  nomenclature proposed in D'Abrusco et al. (2013). Specifically, we refer to the region occupied by $\gamma$-ray emitting blazars in the  3-dimensional \wse\ IR color space as the $locus.$ We refer to its  2-dimensional projection in the [3.4]-[4.6]-[12] $\mu$m color-color diagram as the \wse\ Gamma-ray Strip.

\begin{figure*}[!h]
\begin{center}
\includegraphics[height=12.7cm,width=10.8cm,angle=0]{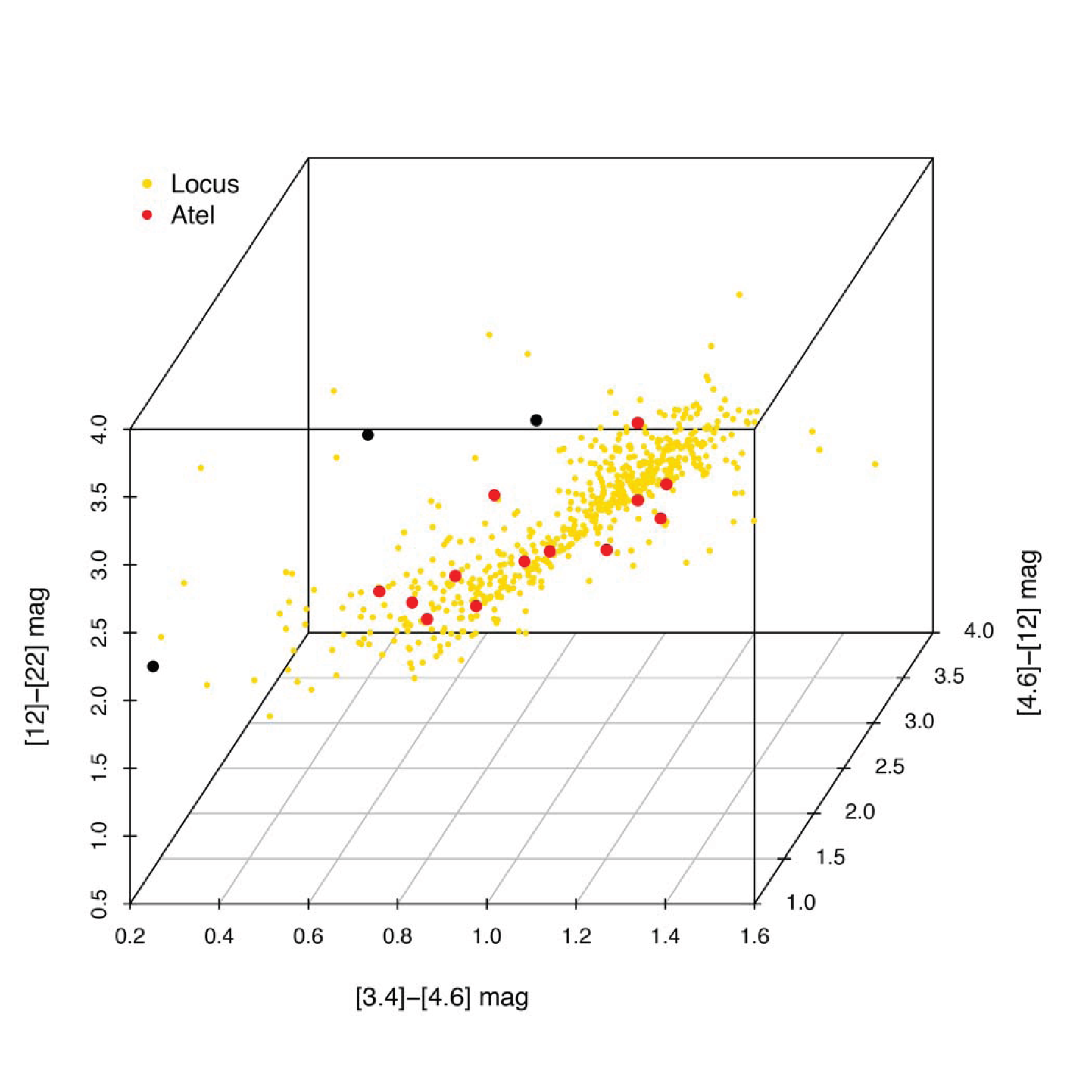}
  \caption{Plot showing the 3-dimensional color-color space of \wse\ . The filled yellow circles indicate the population of known $\gamma\ $-ray blazars\citep[][]{dabrusco12}. The filled red circles indicate the thirteen $gamma$-ray blazar candidates. The filled black circles indicate the three potential blazar candidates (Please see main text for details).}
\label{fig:colors}
\end{center}
\end{figure*}

In this paper, we present our application of the \wse\ association procedure to a sample of 102 sources selected from the literature. Section~\ref{sec:details} presents the details of our sample selection and a source-by-source analysis. Section~\ref{sec:summary} summarizes our findings for this sample and presents our comments on the general efficiency of this method as applied to larger datasets. The IR colors and \wse\ magnitudes of each source can be found in Table~\ref{tab:colors}. An IR color--color plot  comparing the location of these candidate blazars with the known population of  $\gamma$-ray blazars is shown in Figure~\ref{fig:colors}. The angular separations between counterparts are reported in Table~\ref{tab:coordinates} found in Appendix~\ref{app:A}. The complete sample of sources can be found in Appendix~\ref{app:B}.

\section{Searching for Gamma-Ray Blazars}
\label{sec:details}
\subsection{The Association Method}
\label{sec:method}
In this paper, we investigate whether or not transient sources recently reported in The Astronomer's Telegrams (ATels)\footnote{http://www.astronomerstelegram.org/} can be associated to \wse\ counterparts whose IR colors are consistent with those of $\gamma$-ray blazars. Specifically, we searched the entire ATel database looking for sources that showed high-energy non-periodic variability. We also took into account any multifrequency archival observations. We then excluded sources that had already been identified since the ATel publication date. This produced a sample of 102 objects. The complete list of objects in this sample can be found in Appendix~\ref{app:B}. We should note that variability by itself is a common and very powerful method used when hunting for blazars \citep[see e.g.][]{ruan12}.

We then applied our association method to the sources in our sample to determine if there exists a low-energy counterpart showing IR colors consistent with the $locus$ of blazars in the \wse\ color space. The \wse\ survey operated at 3.4, 4.6, 12, and 22 $\mu$m with a typical positional uncertainty of $\sim$1\arcsec \citep[see][for more details]{wright10}.\footnote{http://wise2.ipac.caltech.edu/docs/release/prelim/} 

The identification of candidate blazars was performed using the original association method discussed in D'Abrusco et al. (2013). This method is based on the largest available sample of bona fide blazars, the ROMA BZCat \citep[][]{massaro09,massaro10,massaro11a}, which have been reliably associated to $\gamma$-ray sources listed in the 2FGL catalog. D'Abrusco et al. (2012) found that this sample of $\gamma$-ray emitting blazars is located in a distinct region of the three-dimensional \wse\ color space (see Figure~\ref{fig:colors} and Figure 1 in D'Abrusco et al. 2012), called the locus. This region is almost devoid of non-blazar sources. 

The locus is best modeled in the three-dimensional space generated by the Principal Components (PC) of the color distribution for $\gamma$-ray emitting blazars (see D'Abrusco et al. 2013 for details). The association method, based on 3D modeling of the locus in the PC space, evaluates the compatibility of a generic \wse\ source with the geometric model of the locus in the 3D PC space by comparing their relative positions. The method also takes into account the uncertainties on each \wse\ color and provides a finer classification of candidate blazars (i.e. BZB or BZQ). The performance of the association method, expressed in terms of its purity and completeness as a function of \wse\ colors and sky coordinates, has been evaluated and discussed in D'Abrusco et al. (2013).

Using this methodology, we were able to identify thirteen $\gamma$-ray blazar candidates from the original sample of 102. There are three additional sources whose IR colors are consistent with the blazar population but not specifically with the locus occupied by $\gamma$-ray emitting blazars \citep[see][]{dabrusco12, paper6, massaro13}. However, their peculiar multifrequency properties are consistent with blazar-like sources. 

We searched for additional multifrequency observations of our candidates from the following databases. When looking for radio counterparts we searched the NRAO VLA Sky Survey \citep[NVSS;][]{condon98}, the VLA Faint Images of the Radio Sky at Twenty-Centimeters \citep[FIRST;][]{white97}, Sydney University Molonglo Sky Survey \citep[SUMSS;][]{mauch03} and the The Australia Telescope 20 GHz Survey \citep[AT20G; ][]{murphy10}. NVSS and FIRST both operate at 1.4 GHz with typical positional uncertainties of $\sim$5--7\arcsec\ and $\sim$1--2\arcsec , respectively. SUMSS operates at 843 MHz with a positional uncertainty of $\sim$8\arcsec . Lastly, AT20G operates at 20 GHz with a positional uncertainty of $\sim$1\arcsec .

We were able to identify counterparts in the Two Micron All Sky Survey \citep[2MASS;][]{skrutskie06} by using the previous associations included in the standard \wse\ catalogue \citep[see][]{cutri12}. We then searched for potential optical counterparts, with a particular focus on those with available spectroscopic data. We made use of the Sloan Digital Sky Survey \citep[SDSS DR9; e.g.][]{adelman08, paris12}, which operates in the u,g,r,i and z bands with a typical positional error of $\sim$2\arcsec .

We then searched for high energy counterparts by looking in X-ray survey missions. In the soft X-rays we use the ROSAT all-sky catalogue \citep{voges99}. This allows us to look for detections in the 0.1--2 keV band with a positional uncertainty of $\sim$12\arcsec . We identified potential $\gamma$-ray counterparts by searching the 2FGL \citep[e.g. ][]{ackermann11, nolan12}. We looked for detections at the 95\% level of confidence, which is $\sim$0.1\degr\ across the 30 MeV -- 30 GeV band.

We also made use of the NASA Extragalactic Database (NED) \footnote{\underline{http://ned.ipac.caltech.edu/}} for additional resources and information about previous work on these sources.

\subsection{$\gamma$-Ray Blazar Candidates}
\label{sec:gammablazars}
We found thirteen sources in the transient ATels list that have multi-wavelength characteristics consistent with $\gamma$-ray blazar behavior. They can be associated with a \wse\ counterpart with IR colors that fall within the $locus$ of $\gamma$-ray blazars.

\subsubsection{1RXS J002159.2-514028} 
The unidentified X-ray source 1RXS J002159.2-514028 was identified as a potential BL Lac candidate \citep[][]{mahony10}. The source has several suggested counterparts. It is associated with the radio source SUMSS J002159-514025 \citep[][]{mauch03}. The \swf\ master catalogue indicates that there are also two X-ray sources associated with 1RXS J002159.2-514028. Both have recently been observed by the \swf\ XRT, UVOT, and BAT instruments\footnote{http://www.iasf-palermo.inaf.it/cgi-bin/INAF/pub.cgi?href=activities/bat/index.html} \citep[][]{cusumano10}.

Follow-up analysis was performed using our association procedure on the radio coordinates. Considering an area consistent with the 2\arcsec\ error radius in the SUMSS catalogue, we identify the unique candidate \wse\ source, WISE J002200.08-514024.2 as the IR counterpart to 1RXS J002159.2-514028. Its WISE IR colors are consistent with those of BL Lac objects.

The IR position of WISE J002200.08-514024.2 is consistent (0\arcsec.3 offset) with the NVSS coordinates of  1RXS J002159.2-514028 \citep[][]{condon98}. A 2MASS counterpart, 2MASS 00220011-5140242, is located 0\arcsec.3 from the \wse\ coordinates. This 2MASS source was perviously associated with 1RXS J002159.2-514028 by Haakonsen \& Rutledge (2009).

Within the 2FGL catalogue, 2FGL J0022.3-5141 is the closest $\gamma$-ray source to the IR coordinates. The source lies 3\arcmin.7 from the \wse\ position of 1RXS J002159.2-514028, lying within the uncertainty positional region at the 95\% level of confidence as reported in the 2FGL, indicating the existence of a $\gamma$-ray counterpart to 1RXS J002159.2-514028.

On the basis of its IR colors, coupled with the existence of several multi-wavelength counterparts we conclude that \object{WISE J002200.08-514024.2} is associated with the BL Lac object 1RXS J002159.2-514028, alias SUMSS J002159-514025.

\subsubsection{1RXS J013106.4+612035} 
The X-ray source 1RXS J013106.4+612035 discovered by ROSAT
has been associated with the low energy counterpart  
NVSS J013107+612033 \citep[][]{parades02}. This radio source shows 
a single side relativistic jet \citep[][]{ribo02}.
The optical spectrum is featureless \citep[][]{marti04}.
A detected \textsc{Na I} interstellar absorption feature
suggests that this object could be highly reddened \citep[][]{marti04}.

According to both the 2LAC catalogue and the 2FGL,
the $\gamma$-ray source
2FGL J0131.1+6121 has been associated with 1RXS J013106.4+612035,
and classified as an Active Galactic Nucleus of Uncertain type \citep[e.g. ][]{ackermann11, nolan12}

Searching a region corresponding to the $\gamma$-ray positional uncertainty
at the 95\% level of confidence, we found that the IR source WISE J013107.22+612033.4
is the unique candidate counterpart for 2FGL J0131.1+612. Its IR colors 
are consistent with those of $\gamma$-ray emitting blazars.

The IR position of WISE J013107.22+612033.4 is consistent 
with the radio position of NVSS J013107+612033
lying only 0\arcsec.7 from the NVSS position \citep[][]{condon98}
and within the 7 $\arcsec$ positional uncertainty reported for
1RXS J013106.4+612035 \citep[][]{voges99}.

The source is associated with the 2MASS counterpart 2MASS 01310723+6120334. The association between the X-ray and the 2MASS
sources was also found by Haakonsen \& Rutledge (2009).

On the basis of the peculiar IR colors of \object{WISE J013107.22+612033.4}
coupled with the optical spectrum found in literature
and its X-ray and radio counterparts, we conclude that
the \wse\ source is the unique candidate counterpart associated with the $\gamma$-ray blazar 1RXS J013106.4+612035, alias NVSS J013107+612033.

\subsubsection{IGR J02341+0228} 
The unidentified \inte\ source IGR J02341+0228 has been associated with the quasar QSO B0231+022 \citep[][]{veron06}. We consider the radio source NVSS J023348+022924 as the primary radio counterpart \citep[][]{condon98}. An optical counterpart exists in the Palomar Prime Focus Universal Extragalactic Instrument (PFUEI) CCD survey \citep[][]{schmidt86}.

The radio counterpart is associated to the ROSAT source 1RXS J023349.3+022933. There is no $\gamma$-ray source associated with this NVSS source in the 2LAC catalogue or the 2FGL.

We identify the IR source WISE J023349.19+022925.1 as the unique candidate counterpart to QSO B0231+022 and verify that the IR colors of the \wse\ counterpart are consistent with those of $\gamma$-ray emitting blazars.

The \wse\ source position is positionally consistent (1\arcsec.3 offset) with the position of QSO B0231+022. It is also positionally consistent (3\arcsec.2 offset) with the radio coordinates for NVSS J023348+022924. Additionally, there is a 2MASS counterpart, 2MASS J02334920+0229251, located 0\arcsec.1 from the \wse\ position. We note that this 2MASS counterpart and ROSAT counterpart had been previously associated by Haakonsen \& Rutledge (2009).

On the basis of the distinct IR colors of  WISE J023349.19+022925.1, coupled with the presence of associated multifrequency counterparts found in literature we conclude that the IR source \object{WISE J023349.19+022925.1} is associated with the $\gamma$-ray blazar QSO B0231+022, alias IGR J02341+0228.

\subsubsection{CRATES J0531-4827} 
The \fer\  $\gamma$-ray source 2FGL J0532.0-4826 was discovered in the vicinity of the radio source  CRATES J0531-4827, alias SUMSS J053158-482737,  by Cannon et al. (2010). The object is a known flat-radio spectrum source \citep[][]{healey07}. The authors reported that the gamma-ray emission was indicative of flaring, the type of variability most strongly associated with blazar-like activity. Indeed, this object is a previously known candidate blazar \citep[][]{abdo10a} and is identified as an AGN of uncertain type in the 2FGL catalogue \citep[][]{ackermann11}. The source is also present in the ROSAT catalogue, being associated with the X-ray object 1RXS J053159.9-482751.

The IR colors of the WISE counterpart, WISE J053158.61-482736.0, associated with SUMSS J053158-482737 are consistent with those of $\gamma$-ray emitting blazars. 

The ROSAT source,1RXS J053159.9-482751, associated with the radio position is offset from the \wse\ coordinates by  19\arcsec.7. It is offset from the \fer\ source by 1\arcmin.2. This is still consistent with the \fer\ error circle at the 95\% confidence level. There is no associated 2MASS source for this object.

On the basis of its distinct IR colors coupled with the multifrequency observations we confirm that  \object{WISE J053158.61-482736.0} is the unique IR counterpart of the $\gamma$-ray blazar CRATES J0531-4827, alias SUMSS J053158-482737.
 
 \subsubsection{NVSS J053942+143345} 
The detection of an unidentified \fer\ LAT $\gamma$-ray source, positionally consistent with the radio source TXS 0536+145, alias NVSS J053942+143345, was first reported by Orienti et al. (2012). The source is not  present in the 2FGL catalogue. NVSS J053942+143345 is a known candidate blazar \citep[][]{sowards05} of unknown redshift.

The source has no counterpart in the ROSAT catalogue. The \swf\ catalogue indicates three sources associated with TXS 0536+145, alias NVSS J053942+143345. The most probably counterpart, {\it SWXRT} J053942.4+143346, can be clearly seen in the \swf\ image. It lies 1\arcsec\ from the NVSS position.

Running our association procedure, we verified that the IR colors of the WISE counterpart, WISE J053942.37+143345.5 associated with NVSS J053942+143345, are consistent with those of $\gamma$-ray emitting blazars.

The \wse\ coordinates are positionally consistent (0\arcsec.3 offset) with the VLBI coordinates and are separated from the original \fer\ source by 1\arcmin.8. There is a single 2MASS source, 2MASS J05394237+1433455, which is 0\arcsec.1 offset from the \wse\ coordinates. 

On the basis of the unique IR colors of \object{WISE J053942.37+143345.5}, coupled with the multifrequency observations available in the literature, we conclude that it is the IR counterpart of the $\gamma$-ray blazar NVSSJ053942+143345.

\subsubsection{1RXS J072259.5--073131} 
The unidentified  ROSAT object, 1RXS J072259.5--073131, has been identified as the high-energy counterpart to the radio source NVSS J072259-073135 \citep{parades02}. The radio structure shows a curiously bent one-sided relativistic radio jet \citep[][]{ribo02}. The optical spectrum reported by Matri et al. (2004) is devoid of any strong features. The source does not have a counterpart in the 2FGL catalogue.

Running our association procedure, we verified that the IR colors of the WISE counterpart,WISE J072259.68--073135.0 associated with 1RXS J072259.5--073131, are consistent with those of $\gamma$-ray emitting blazars. 

The \wse\ coordinates are located 1\arcsec\ from the NVSS position and 4\arcsec.5 from the ROSAT coordinates. The \wse\ catalogue indicates a 2MASS counterpart, 2MASS 07225968--0731347, located 0\arcsec.3 from the \wse\ coordinates. We note, also in this case, the previously identified association between the 2MASS and ROSAT sources by Haakonsen \& Rutledge (2009). 

On the basis of the unique IR colors, coupled with the multifrequency observations available in the literature, we conclude that the source \object{WISE J072259.68--073135.0} is the IR counterpart to the BL Lac object 1RXS J072259.5--073131, alias NVSS J072259-073135.

\subsubsection{CSS100408:073324+365005} 
The Catalina Real-time Transient Survey (CRTS, see e.g., Drake et al. 2012)\footnote{http://crts.caltech.edu/} reported a radio flare by CSS100408:073324+365005, alias GB6 B0730+3657,  in April 2010 \citep[][]{mahabal10}. The flare coordinates are associated with the radio source NVSS J073324+365004. The peak apparent magnitude of the flare was $\sim$17.7 magnitude in the V--band. Follow up optical analysis by Djorgovski et al. (2010) using the optical spectrum from Keck showed an {\tt [O II]} 3737 emission line and 4000 $\AA$ break indicating a spectroscopic redshift of {\it z}=1.235 \citep[][]{djorg10}.

The source does not appear to have any known high-energy counterparts. There are no \swf\ sources within 20\arcmin\ of the radio position. We searched the 2FGL catalogue finding the nearest \fer\ source to be $\sim$4\degr\ from the NVSS coordinates.

We verified that the IR colors of the WISE counterpart,WISE J073324.38+365004.5, associated with NVSS J073324+365004 are consistent with those of $\gamma$-ray emitting blazars. 

 The \wse\ source is positionally consistent (1\arcsec.1 offset) with the NVSS coordinates, indicating a good association with the radio counterpart. There is no identified 2MASS counterpart to the \wse\ source. 

On the basis of the peculiar IR colors, coupled with the multifrequency observations available in the literature we conclude that the source \object{WISE J073324.38+365004.5} is the IR counterpart of the $\gamma$-ray blazar NVSS J073324+365004.

\subsubsection{B3 0819+408} 
The radio source B3 0819+408, alias NVSS J082257+ 404149,  has been previously classified as a quasar and is a candidate blazar \citep[see e.g. ][]{young09, miller11}. It is identified as a flat-spectrum radio source. The source has an optical counterpart in the SDSS Data Release 9 \citep[][]{paris12}. The source SDSS J082257.55+404149.8 is identified as a broad-line QSO.

The source has an X-ray counterpart in the XMM source catalogue, 2XMM J082257.6+ 404149. Searching within the 2FGL catalogue, we note that B3 0819+408 is a known counterpart to the \fer\ source 2FGL J0823.0+4041. It is classified as an AGN of uncertain type.

We verified that the IR colors of the WISE counterpart associated with NVSS J082257+ 404149 are consistent with those of the $\gamma$-ray emitting blazars. Searching a patch of sky consistent with the 1\arcsec\ NVSS error circle we identify the \wse\ source, WISE J082257.55+404149.8, as the unique candidate IR counterpart to B3 0819+408.

The \wse\ position is positionally consistent (0\arcsec.6 offset) with the VLBA coordinates. It is offset from the XMM coordinates by 0\arcsec.6. The separation between the \fer\ counterpart and \wse\ position is 1\arcsec.4.  The \wse\ catalogue does not indicate the presence of a 2MASS counterpart.

On the basis of the unique IR colors, coupled with the multifrequency observations available in the literature we conclude that the source \object{WISE J082257.55+404149.8} is the unique IR counterpart to the $\gamma$-ray blazar B3 0819+408, alias NVSS J082257+404149.

\subsubsection{1RXS J130737.8-425940} 
The unidentified X-ray source 1RXS J130737.8-425940, has been associated with the source 2FGL J1307.5-4300, identified as an AGN of uncertain type in 2FGL. The radio counterpart is SUMSS J130737-425940.

1RXS J130737.8-425940 is detected in several high-energy catalogues. The \swf\ catalogue indicates that there is one X-ray source associated 1RXS J130737.8-425940. The $\gamma$-ray source 2FGL J1307.5-4300, lies 0\arcmin.8 from 1RXS J130737.8-425940. This is within the 3\arcmin\ positional uncertainty region at the 95\% level of confidence for \fer\ and indicates a $\gamma$-ray counterpart to 1RXS J130737.8-425940.

 Searching an area corresponding to the 10\arcsec\ ROSAT error circle we identify the IR source WISE J130737.98-425938.9 as the unique counterpart to 1RXS J130737.8-425940. We verified the counterpart has IR colors consistent with $\gamma$-ray emitting blazars.

The \wse\ source is offset from the SUMSS coordinates by 2\arcsec.9, this is just outside the 2\arcsec\ error circle. There is a 2MASS counterpart, 2MASS 13073797-4259389, located 0\arcsec.2 from the \wse\ coordinates, that has been previously associated with 1RXS J130737.8-425940 by Haakonsen et al. (2009).

On the basis of the unique IR colors, coupled with the multifrequency observations available in the literature we conclude that the source \object{WISE J130737.98-425938.9} is the unique IR counterpart to the blazar 1RXS J130737.8-425940, alias the $\gamma$-ray source 2FGL J1307.5-4300.

\subsubsection{PMN J1532-1319} 
The detection of a $\gamma$-ray flare by \fer\ LAT, associated with the radio source TXS 1530-131, alias PMNJ1532-1319, was first reported by Gasparrini et al. (2011). The radio source has a known optical counterpart  in the Candidate Gamma Ray Blazar Survey (CGRaBS), with an apparent magnitude of 22.2 in the R--band \citep[][]{healey08}. The object is identified as a flat-spectrum radio source in NED.

The \swf\ catalog indicates a potential counterpart, {\it SWXRT} J153245.7-131909,  5\arcsec.1 from the radio position. The catalogue indicates the source has been previously associated with PMN J1532-1319.

We applied our association procedure, searching for potential IR candidate counterparts to the \fer\ flare. We were able to identify two \wse\ sources positionally consistent with the \fer\ LAT coordinates and verified that both of these candidate sources have IR colors characteristic of $\gamma$-ray detected blazars.

WISE J153245.37-131910.0 is positionally consistent (1\arcsec\ offset) with the VLBA coordinates. The other candidate, WISE J153355.64-132607.9, is offset from the radio coordinates by 0\degr.3. Thus WISE J153245.37-131910.0 is the more probable IR candidate counterpart to PMN J1532-1319. There is no 2MASS counterpart.

On the basis of the unique IR colors, coupled with the multifrequency observations available in the literature, we conclude that \object{WISE J153245.37-131910.0}, is the unique IR counterpart of the $\gamma$-ray blazar PMN J1532-1319.

 \subsubsection{NVSS J191320-363019} 
The detection of a gamma-ray flare by \fer\ LAT, positionally consistent with the radio source VCS4 J1913-3630, alias NVSS J191320-363019, was first reported by Donato et al. (2010). The detection of flaring activity by \fer\ is consistent with blazar-like activity. The source is identified in NED as a flat spectrum radio source of unknown redshift.
 
 Searching the high-energy catalogues we find no counterparts in the \swf\ master catalogue within a 5\arcmin\ radius. Likewise, the nearest ROSAT source is 4\arcmin.8 away. There is no associated $\gamma$-ray counterpart in the 2FGL.

We verified, via our association procedure, that the IR colors of the WISE counterpart, WISE J191320.89-363019.5, associated with NVSS J191320-363019 are consistent with those of the $\gamma$-ray emitting blazars.

The \wse\ coordinates are strongly positionally consistent (0\arcsec.5 offset) with the NVSS coordinates \citep[][]{condon98}. There is no associated 2MASS source for this object.

On the basis of its distinctive IR colors, coupled with the multifrequency observations available in the literature, we conclude that  \object{WISE J191320.89-363019.5}, is the unique IR counterpart of the $\gamma$-ray blazar, NVSS J191320-363019.

\subsubsection{NVSS J222329+010226} 
The unidentified radio source NVSS J222329+010226 does not appear in any other radio catalogues nor is there an associated counterpart in the SDSS DR9 catalogue.

We investigated the high-energy catalogues for a counterpart. The nearest \swf\ sources are $\sim$ 5\arcmin\ from the NVSS coordinates. There are no nearby ROSAT or \chn\ sources. The object does have a $\gamma$-ray counterpart. The source, 2FGL J2223.4+0104, is located 1\arcmin.94 from the radio coordinates and has been previously associated with NVSS J222329+010226.

We verified that the IR colors of WISE J222329.56+010226.7, associated with NVSS J222329+010226, are consistent with those of the $\gamma$-ray emitting blazars. The \wse\ coordinates are positionally consistent (0\arcsec.4 offset) with the NVSS coordinates. The \wse\ source is offset from the \fer\ source by 1\arcmin.9. This is within the positional uncertainty region at the 95\% level of confidence reported in the 2FGL. The \wse\ catalogue does not indicate the existence of a 2MASS counterpart.

On the basis of its unique IR colors, coupled with the multifrequency observations available in the literature we conclude that \object{WISE J222329.56+010226.7}, is the unique IR counterpart of the $\gamma$-ray blazar, NVSS J222329+010226.

\subsubsection{PMN J2258-8246} 
The unidentified radio source PMN J2258-8246, alias SUMSS J225759-824650, has been identified as a potential $\gamma$-ray blazar candidate \citep[][]{abdo10a}. The source does not appear in any other radio or optical surveys. 

The object does appear in several high-energy catalogues. The \swf\ catalog indicates that PMN J2258-8246 has been observed by the \swf\/XRT, UVOT, and BAT instruments.  The \swf\ position is consistent (1\arcsec.1 offset) with the SUMSS coordinates. 

Within the 2FGL catalogue, the closest $\gamma$-ray source, 2FGL J2259.0-8254, lies 7\arcmin.8 from the SUMSS position. This is on the edge of the  7\arcmin\ positional uncertainty region at the 95\% level of confidence, indicating a possible $\gamma$-ray counterpart to PMN J2258-8246. 

We verified that the IR colors of the WISE counterpart, WISE J225759.01--824652.5 associated with PMN J2258-8246, are consistent with those of the $\gamma$-ray emitting blazars.

The \wse\ position is consistent (2\arcsec\ offset) with the SUMSS coordinates. They are also consistent with the \fer\ position, being offset by 7\arcmin.7. The \wse\ catalogue indicates a 2MASS counterpart, 2MASS J22575948-8246531, located 1\arcsec.1 from the \wse\ coordinates.

On the basis of its peculiar IR colors, coupled with the multifrequency observations available in the literature, we conclude that \object{WISE J225759.01--824652.5},is the unique IR counterpart to the $\gamma$-ray blazar, PMN J2258-8246 alias 

\subsection{Alternate Blazar-Like Sources}
\label{sec:altsources}
Among our sample we found three sources that display multi-wavelength behavior indicative of blazars, but whose \wse\ counterparts feature IR colors only consistent with those of the general blazar population, but not specifically with the $locus$ of $\gamma$-ray blazars. Here we discuss these sources in detail.

\subsubsection{1RXS J001442.2+580201} 
The unidentified X-ray source 1RXS J001442.2+580201 has been identified as the high-energy counterpart to the radio source NVSS J001441+580202 \citep[][]{parades02}. The radio structure clearly shows the presence of double-sided jets \citep[][]{ribo02}. The optical spectrum is completely featureless. The source appears to be optically variable \citep[][]{marti04}.

We verified that the IR colors of the \wse\ counterpart associated with 1RXS J001442.2+ 580201 are consistent with those of the blazar population. However, they are not consistent with the {\it locus} occupied by $\gamma$-ray emitting blazars. We identify the source WISE J001442.11+580201.2 as the unique IR counterpart to 1RXS J001442.2+580201. The IR colors are similar to BZB objects \citep[][]{massaro12a, dabrusco12, paper6}.

Its position is consistent with the radio counterpart, lying only 0\arcsec.1 from the NVSS position \citep[][]{condon98}.  This is well within the 3\arcsec\ error radius as reported by Parades et al. (2002). 

The \wse\ catalogue lists no associated 2MASS source for this object. The nearest object in 2FGL lies 3\degr.3 from the \wse\ coordinates. 

On the basis of our IR color analysis, combined with the optical spectrum found in literature and the presence of radio and X-ray counterparts,  we conclude that the \wse\ source \object{WISE J001442.11+580201.2}, is associated with the BL Lac object 1RXS J001442.2+580201, alias NVSS J001441+580202.

\subsubsection{1RXS J032521.8-563543} 
The unidentified X-ray source 1RXS J032521.8-563543 has been previously identified as an AGN of uncertain type in recent surveys \citep[][]{abdo10a, mahony10, ackermann11}. The source has been associated with the low energy counterpart SUMSS J032523-563545 \citep[][]{mahony10}. It has also been associated with the known flat-spectrum radio source CRATES J032523.52-563544.3 \citep[][]{healey07}.

1RXS J032521.8-563543 is also detected in other high-energy catalogues. The \swf\ catalogue indicates that there is a unique X-ray source associated with the radio counterpart of 1RXS J032521.8-563543. The closest gamma-ray source, reported in 2FGL, 2FGL J0325.1-5635, lies 1\arcmin.5 from the ROSAT position of 1RXS J032521.8-563543. This source is indicated as an AGN of uncertain type and a known $\gamma$-ray counterpart to 1RXS J032521.8-563543. 

The X-ray source has been associated with the 2MASSX Extended Source 2MASSX 03252346-5635443. There is also a point source 2MASS counterpart, 2MASS 03252353-5635446. These associations make it easy to identify the \wse\ counterpart as WISE J032523.51-563544.6. Its IR colors are consistent with those of the blazar population but not specifically with the {\it locus} occupied by $\gamma$-ray emitting blazars.

The \wse\ source is positionally consistent (0\arcsec.5 offset) with the 2MASSX coordinates. It is separated from the X-ray coordinates by 14\arcsec\ , just outside the 10\arcsec\ ROSAT 1$\sigma$ error circle. The offset between the \wse\ and \fer\ coordinates is 1\arcmin.8. This is within the positional uncertainty region at the 95\% level of confidence for \fer\ data.

On the basis of the unique IR colors of \object{WISE J032523.51-563544.6}, coupled with the presence of associated multifrequency counterparts found in literature we conclude that the source is associated with the $\gamma$-ray blazar 2FGL J0325.1-5635, alias 1RXS J032521.8-563543 and SUMSS J032523-563545.

\subsubsection{\fer\ J1717-5156} 
The discovery of the \fer\ LAT transient gamma-ray source \fer\ J1717-5156 was first reported by Schinzel et al. (2012). The authors indicated the unidentified radio source PMN J1717-5155 is a potential counterpart to the \fer\ transient. Using flux density measurements from Wright et al. (1994) and Murphy et al. (2007) they found the source to have a nearly flat radio spectrum with $\alpha$ = 0.5 \citep[][]{schinzel12}. Spectroscopic observations identified broad emissions lines for \textsc{C [III]} ($\lambda = 1909\AA$) and \textsc{Mg II} ($\lambda = 2798\AA$) at a redshift of {\it z} = 1.158, consistent with the sources blazar-like nature \citep[][]{chomiuk13}.

Follow-up observations were performed by \swf\ XRT/UVOT. The XRT observations identified a likely x-ray counterpart to the $\gamma$-ray source. Three other potential counterparts in the \fer\ error circle at the 95\% level of confidence were also identified. The UVOT observations revealed the presence of  potential optical variability. Details of these observations can be found in Lu et al. (2012) and Cheung et al. (2012). The source does not appear in the 2FGL.

Our association procedure found that the IR colors of the WISE counterpart, WISE J171734.65-515532.0 associated with \fer\ J1717-5156, are consistent with those of the general blazar population but not the distinct $locus$ of $\gamma$-ray emitting blazars. 

The \wse\ source is found to be positionally consistent (13\arcsec\ offset) with the PMN coordinates. It is located 1\arcsec.8 from the primary \swf\ XRT counterpart \citep[][]{lu12}. There is no 2MASS counterpart.

On the basis of the distinctive IR colors, coupled with the multifrequency observations available in the literature we conclude that \object{WISE J171734.65-515532.0}, is the unique IR candidate counterpart of the $\gamma$-ray blazar \fer\ J1717-5156, alias PMN J1717-5155.

\section{Summary and Conclusions}
\label{sec:summary}
We have presented a detailed discussion of the multi-wavelength counterparts for sixteen candidate blazars. We were able to identify 13 $\gamma$-ray blazar candidates associated via our \wse\ procedure. We also identified three additional blazar-like sources. This sub-sample was selected from a larger sample of 102 variable sources. Sources were selected on the basis of variability (a key blazar feature) as reported in the Astronomer's Telegrams and literature. Candidates were selected from this sample by identifying an IR counterpart with colors consistent with the $locus$ of $\gamma$-ray emitting blazars in the \wse\ color-color space.
   
The general properties of the sample can be summed up as follows:

\begin{enumerate}
\item Thirteen of the sixteen candidate sources have IR colors consistent with the population of $\gamma$-ray emitting blazars (see Figure 1). The remaining three have IR colors consistent with the blazar population, but not specifically the $locus$ of $\gamma$-ray blazars. 
\item Only four of the sources in our sub-sample have been previously identified as candidate active galaxies, specifically QSOs. A further four have been identified as galaxies. This  is in agreement with nearby BZB objects since in many cases we can only see their host galaxy via optical spectroscopy. Thus, we have been able to confirm the nature of these objects as well as provide a refined candidate classification for the remaining sources.
\item The existence of a radio counterpart is ubiquitous in the sample. All sixteen of the sources can be associated with strong radio emission detected in existing surveys. Blazars are, by definition, radio-loud objects so the fact that all of our candidates have a radio counterpart is encouraging.
\item Five of the sixteen sources can be associated with an optical counterpart. Of those five, three have available spectroscopic data. The available spectra display features consistent with those expected for blazars.
\item Eleven of the sixteen sources can be associated with an X-ray counterpart. The X-ray sources are almost universally from all-sky surveys like ROSAT or \swf\ pointed observations. We expect blazars to emit across the whole electromagnetic spectrum, and in particular they are well known X-ray sources. This is confirmed by our findings.
\item We were able to identify potential \fer\ $\gamma$-ray counterparts in the 2FGL for exactly half of the sub-sample (8 out of 16). Given that our sources were selected due to having IR colors consistent with $\gamma$-ray blazars, we expect the other nine sources to have currently uncatalogued $\gamma$-ray counterparts.
\end{enumerate}

This association method can only be used to produce a list of $\gamma$-ray blazar {\em candidates}. Follow-up spectroscopic observations must be made in order to confirm the blazar nature of these sources. The presence and strength of optical lines is one of the primary observational necessities to both confirm the source as a blazar and assign a specific classification (e.g. BZB vs. BZQ). 

We were able to identify thirteen new candidate $\gamma$-ray blazars and three blazar-like sources, with only four previously known, indicating that the IR colors is potentially both predictive and efficient to recognize non-thermal emission. This is especially valuable in the era of large time-resolved optical surveys such as PanSTARRS and, even more so, LSST.

\begin{acknowledgements}
We thank our anonymous referee for many helpful comments which greatly improved this paper. The work is supported by the NASA grants NNX10AD50G, NNH09ZDA001N and NNX10AD68G. PSC is grateful for support by the NSF REU and DOD ASSURE programs
under NSF grant no. 0754568 and by the Smithsonian Institution. R. D'Abrusco gratefully acknowledges the financial 
support of the US Virtual Astronomical Observatory, which is sponsored by the
National Science Foundation and the National Aeronautics and Space Administration.
F. Massaro acknowledges the Foundation BLANCEFLOR Boncompagni-Ludovisi, n'ee Bildt
for the grant awarded him in 2010 to support his research.
The work by G. Tosti is supported by the ASI/INAF contract I/005/12/0.
TOPCAT\footnote{\underline{http://www.star.bris.ac.uk/$\sim$mbt/topcat/}} 
\cite{Taylor2005} and SAOImage DS9 were used extensively in this work 
for the preparation and manipulation of the tabular data and the images.
This research has made use of data obtained from the High Energy Astrophysics Science Archive
Research Center (HEASARC) provided by NASA's Goddard
Space Flight Center; the SIMBAD database operated at CDS,
Strasbourg, France; the NASA/IPAC Extragalactic Database
(NED) operated by the Jet Propulsion Laboratory, California
Institute of Technology, under contract with the National Aeronautics and Space Administration.
Part of this work is based on archival data, software or on-line services provided by the ASI Science Data Center.
This publication makes use of data products from the Wide-field Infrared Survey Explorer, 
which is a joint project of the University of California, Los Angeles, and 
the Jet Propulsion Laboratory/California Institute of Technology, 
funded by the National Aeronautics and Space Administration.
\end{acknowledgements}

  \begin{table*}[!h]
\caption{\wse\ Colors and Known Counterparts of Candidate Blazars.\label{tab:colors}}
\centering
\tiny
\begin{tabular}{|llccclc|}
\hline
  Source &  \wse\  & [3.4]-[4.6] & [4.6]-[12] & [12]-[22] & notes & z \\
  name   &  name   &     mag     &    mag     &   mag     &       &   \\
\hline
\noalign{\smallskip}
  1RXS J002159.2-514028  &  J002200.08-514024.2  &  0.81(0.03) &  2.23(0.04) &  1.87(0.14) &  S,M,6,X,v  &  ?      \\
  1RXS J013106.4+612035  &  J013107.22+612033.4  &  0.64(0.04) &  1.88(0.09) &  2.21(0.26) &  N,M,X,BL   &  ?      \\
  IGR J02341+0228        &  J023349.19+022925.1  &  1.08(0.03) &  2.39(0.03) &  2.18(0.05) &  N,M,X,QSO  &  0.321  \\
  PMNJ0531-4827          &  J053158.61-482736.0  &  0.89(0.03) &  2.46(0.03) &  2.06(0.03) &  S,A,X,v    &  ?      \\ 
  NVSSJ053942+143345     &  J053942.37+143345.5  &  1.05(0.06) &  3.13(0.09) &  2.62(0.21) &  N,M        &  2.69?  \\
  1RXS J072259.5-073131  &  J072259.68-073135.0  &  0.71(0.04) &  1.93(0.06) &  2.10(0.24) &  N,M,X      &  ?      \\
  CSS100408:073324+365005&  J073324.38+365004.5  &  0.95(0.04) &  2.41(0.05) &  2.16(0.15) &  N,v        &  1.235      \\
  B3 0819+408            &  J082257.55+404149.8  &  1.15(0.04) &  2.91(0.04) &  2.32(0.07) &  N,M,s,x    &  0.865? \\
  1RXS J130737.8-425940  &  J130737.98-425938.9  &  0.73(0.03) &  2.04(0.03) &  1.90(0.08) &  M,X,v      &  ?      \\
  PMN J1532-1319         &  J153245.37-131910.0  &  1.11(0.04) &  2.69(0.06) &  2.35(0.16) &  N,A        &  ?      \\
  PMNJ1913-3630          &  J191320.89-363019.5  &  1.13(0.06) &  2.93(0.11) &  2.06(0.38) &  N,S,A      &  ?      \\
  NVSS J222329+010226    &  J222329.56+010226.7  &  0.83(0.05) &  2.43(0.13) &  2.56(0.33) &  N,s        &  0.21?  \\
  PMN J2258-8246         &  J225759.01-824652.5  &  0.78(0.03) &  2.16(0.05) &  2.15(0.16) &  S,M        &  ?      \\ 
\noalign{\smallskip}
\hline
  1RXS J001442.2+580201  &  J001442.11+580201.2  &  0.45(0.05) &  3.12(0.07) &  2.55(0.16) &  N,X,v,BL   &  ?      \\
  1RXS J032521.8-563543  &  J032523.51-563544.6  &  0.18(0.03) &  1.52(0.05) &  1.90(0.18) &  S,M,6,X    &  0.06?  \\ 
  FERMI J1717-5156       &  J171734.65-515532.0  &  0.81(0.07) &  3.23(0.09) &  2.58(0.19) &             &  1.158      \\
\noalign{\smallskip}
\hline
\end{tabular}\\
\begin{flushleft}
Col. (1) 2FGL name. \\
Col. (2) \wse\ name. \\
Cols. (3,4,5) Infrared colors from \wse. Values in parentheses are 1$\sigma$ uncertainties. \\
Col. (6) Notes: N = NVSS, F = FIRST, S = SUMSS, A=AT20G, M = 2MASS, s = SDSS dr9, 6 = 6dFG, x = \xmm\ or \\
 \chn, X = ROSAT;  QSO  = quasar, Sy = Seyfert, LNR = LINER, BL = BL Lac; 
v = variability in \wse\ (var\_flag $>$ 5 in at \\
 least one band). \\
Col. (10) Redshift: ? = unknown, number? = uncertain. 
\end{flushleft}
\end{table*}

\clearpage

\appendix
\section{Detailed Listing of Counterparts}
\label{app:A}
\begin{table*}[h!]
\centering
\tiny
\caption{Details of Candidate Blazar Counterparts. \label{tab:coordinates}}
\resizebox{!}{0.32\paperheight}{
\begin{tabular}{|lllll|}
\hline
NAME & RADIO                     &   INFRARED                   &   OPTICAL                   &   X-rays                   \\
          & &                           & &                          \\
\hline
\noalign{\smallskip} 
J0014+5802 & NVSS J001441+580202  & WISE J001442.11+580201.2      &                      & 1RXS J001442.2+580201 \\
Coordinates (J2000) & 00:14:42.1, +58:02:01.32  & 00:14:42.1, +58:02:01.2  &  & 00:14:42.2, +58:02:01.5  \\
Positional uncertainty & 3 & 0.14  &  & 9 \\
Angular Separations from IR position  & 0.1                & 0.0              &                  & 0.7             \\
\hline

\noalign{\smallskip} 
 J0022-5140 & SUMSS J002159-514025  & WISE J002200.08-514024.2      &                    & 1RXS J002159.2-514028 \\
Coordinates (J2000)  & 00:22:00.06, -51:40:24.4  & 00:22:00.08, -51:40:24.2  &   & 00:21:59.20, -51:40:28.5  \\
Positional uncertainty  & 1.50 & 0.08  & & 19 \\
Angular Separations from IR position & 0.3               & 0.0              &                      & 9.3              \\
\hline

\noalign{\smallskip} 
 J0131+6120 & NVSS J013107+612033  & WISE J013107.22+612033.4,      & & 1RXS J013106.4+612035 \\
Coordinates (J2000) & 01:31:07.16, +61:20:33  & 01:31:07.23, +61:20:33.4  &  & 01:31:06.40, +61:20:35  \\
Positional uncertainty & 1.00 & 0.11 &  & 7 \\
Angular Separations from IR position & 0.7                & 0.0              &                     & 6.2              \\
\hline

\noalign{\smallskip} 
 J0233+0229 & NVSS J023348+022924  & WISE J023349.19+022925.1 &  & 1RXS J023349.3+022933 \\
Coordinates (J2000)  & 02:33:48.99, +02:29:24.6  & 02:33:49.2, +02:29:25.2  & &02:33:49.3 +02:29:33  \\
Positional uncertainty & 1.50 & 0.08  &  & 10 \\
Angular Separations from IR position & 3.2                & 0.0              &       & 8              \\
\hline
\noalign{\smallskip} 
 J0325-5635 & SUMSS J032523-563545  & WISE J032523.51-563544.6      & & 1RXS J032521.8-563543 \\
Coordinates (J2000) & 03:25:23.39, -56:35:45.5  & 03:25:23.52, -56:35:44.6  & & 03:25:21.8, -56:35:43.5  \\
Positional uncertainty & 2.10 & 0.08  & & 10 \\
Angular Separations from IR position & 1.3                & 0.0              &  & 14.2              \\
\hline
\noalign{\smallskip} 
 J0531-4827 & CRATES J0531-4827  & WISE J053158.61-482736.0      & & 1RXS J053159.9-482751 \\
Coordinates (J2000)  & 05:31:58.61, -48:27:35.9  & 05:31:58.61, +00:00:00.0  & & 05:31:59.90, -48:27:51  \\
Positional uncertainty & 1.40 & 0.06  &  & 30 \\
Angular Separations from IR position & 0.1                & 0.0              & & 19.7              \\
\hline
\noalign{\smallskip} 
 J0539+1433 & NVSS J053942+143345  & WISE J053942.37+143345.5      & & SWXRTJ053942.4+143346 \\
Coordinates (J2000) & 05:39:42.35, +14:33:45.5  & 05:39:42.37, +14:33:45.5  & & 05:39:42.426 +14:33:45.69  \\
Positional uncertainty & 0.6 & 0.16  &  & 5 \\
Angular Separations from IR position & 0.3                & 0.0              &  & 1              \\
\hline
\noalign{\smallskip} 
 J0722-0731 & NVSS J072259-073135  & WISE J072259.68--073135.0    & & 1RXS J072259.5--073131 \\
Coordinates (J2000)  & 07:22:59.75, -07:31:35.3  & 07:22:55.28, +07:31:35.0  &  & 07:22:59.5, -07:31:31.5  \\
Positional uncertainty & 0.60 & 0.50  &  & 8.00 \\
Angular Separations from IR position & 1.0                & 0.0              & & 4.5              \\
\hline
\noalign{\smallskip} 
 J0733+3650 & NVSS J073324+365004  &WISE J073324.38+365004.5      &  & \\
Coordinates (J2000)  & 07:33:24.48, +36:50:04.7  & 07:33:24.38, +36:50:04.5  & &  \\
Positional uncertainty & 0.70 & 0.90  & & \\
Angular Separations from IR position & 1.1                & 0.0              &        &   \\
\hline
\noalign{\smallskip} 
 J0822+4041 & NVSS J082257+404149  & WISE J082257.55+404149.8   & SDSS J082257.55+404149.8 & 2XMM J082257.6+404149 \\
Coordinates (J2000)  & 08:22:57.50, +40:41:49.7  & 08:22:57.55, +40:41:49.8  & 08:22:57.55, +40:41:49.8  & 08:22:57.50, +40:41:49.8  \\
Positional uncertainty & 0.60 & 0.09  & 2 & 5 \\
Angular Separations from IR position & 0.6               & 0.0              & 0.1                    & 0.6              \\
\hline
\noalign{\smallskip} 
 J1307-4259 & SUMSS J130737-425940  & WISE J130737.98-425938.9      & & 1RXS J130737.8-425940 \\
Coordinates (J2000)  & 13:07:37.74 -42:59:40.1 & 13:07:37.98, -42:59:38.9  & & 13:07:37.8, -42:59:40.5  \\
Positional uncertainty & 2.0 & 0.07  &  & 10.00 \\
Angular Separations from IR position &  2.9         & 0.0              &                 & 2.6              \\
\hline
\noalign{\smallskip} 
 J1532-1319 & NVSS J153245-131909  & WISE J153245.37-131910.0      & CGRaBS J1532-1319 & SWXRTJ153245.7-131909 \\
Coordinates (J2000)  & 15:32:45.34, -13:19:09.2  & 15:32:45.37, -13:19:10.0  & 15:32:45.37, -13:19:10.0  & 15:32:45.684 -13:19:09.05  \\
Positional uncertainty & 0.60 & 0.10  & 0.25 & 5.00 \\
Angular Separations from IR position & 1               & 0.0             & 0.1                    & 4.5              \\
\hline
\noalign{\smallskip} 
 J1717-5155 & PMN J1717-5155  & WISE J171734.65-515532.0       &  & \\
Coordinates (J2000)  & 17:17:35.20, -51:55:20.0  & 17:17:34.65, -51:55:32.1  & &  \\
Positional uncertainty & 126.00 & 0.19  &  &  \\
Angular Separations from IR position &13                & 0.0              &   &             \\
\hline
\noalign{\smallskip} 
 J1913-3630 & NVSS J191320-363019  & WISE J191320.89-363019.5     &    &  \\
Coordinates (J2000)  & 19:13:20.87, -36:30:19.9  & 19:13:20.89, -36:30:19.5  &  &   \\
Positional uncertainty & 0.60 & 0.08  & &  \\
Angular Separations from IR position & 0.5                & 0.0              &    &       \\
\hline
\noalign{\smallskip} 
 J2223+0102 & NVSS J222329+010226  & WISE J222329.56+010226.7      &    &  \\
Coordinates (J2000)  & 22:23:29.59, +01:02:26.7  &22:23:29.56, +01:02:26.7  &   &   \\
Positional uncertainty & 2.90 & 0.13  & & \\
Angular Separations from IR position & 0.4                & 0.0              &  &              \\
\hline
\noalign{\smallskip} 
 J2257-8246 &  SUMSS J225759-824650 & WISE J225759.01-824652.5      & & \\
Coordinates (J2000)  & 22:57:59.36, -82:46:50.7  & 22:57:59.01, -82:46:52.5  & &  \\
Positional uncertainty & 1.70 & 0.09  & & \\
Angular Separations from IR position & 2                & 0.0              & &              \\
\hline
\end{tabular}}\\
\begin{flushleft}
Listing of counterparts for each of the sixteen sources. Coordinates are in the J2000 epoch as reported by their respective catalogues. Positional uncertainties and angular separations are given in arcseconds.
\end{flushleft}
\end{table*}

\clearpage
\section{Complete Listing of Sources in Sample}
\label{app:B}

\begin{center}
 \begin{longtable}{lrrr}
 \caption{Complete Sample of Sources \label{tab:sample}} \\
 \hline
    Source Name & RA (Deg) & Dec (Deg) & ATel\# \\
    \hline
    1RXS J002159.2-514028 & 5.50  & -51.67 & M10  \\
    FERMI J0052+1110 & 13.09 & 11.16 & 3904 \\
    IGR J02115-4407 & 32.87 & -44.13 & 3078 \\
    IGR J02341+0228 & 38.52 & 2.46  & 4102 \\
    1RXS J032521.8-563543 & 51.04 & -56.77 & M10  \\
    1RXS J042201.0+485610 & 65.50 & 48.94 & 3382 \\
    VER J0521+211 & 80.48 & 21.19 & 2260 \\
    IGR J05255-0711 & 81.38 & -7.19 & 2731 \\
    PMN J0531-4827 & 83.00 & -48.46 & 2907 \\
    NVSS J053942+143345 & 84.93 & 14.56 & 3999 \\
    XMMSL1 J063045.9-603110 & 97.69 & -60.52 & 3821 \\
    CSS100408:073324+365005 & 113.35 & 36.84 & 2652 \\
    IGR J08190-3835 & 124.76 & -38.58 & 2975 \\
    B3 0819+408 & 124.90 & 40.86 & M11  \\
    PMN J0822-4308 & 125.69 & -43.14 &  \\
    XRT 1 & 135.53 & -46.38 & 3992 \\
    XRT 2 & 135.56 & -46.51 & 3992 \\
    XRT 3 & 135.69 & -46.30 & 3992 \\
    GRO J0902-35 & 136.20 & -35.25 & 1771 \\
    IGR J10043-8702 & 151.08 & -87.03 & 865 \\
    PMN J1038-5311 & 159.67 & -53.20 & 3978 \\
    IGR J10500-6410 & 162.50 & -64.17 & 865 \\
    IGR J11098-6457 & 167.44 & -64.95 & 1538 \\
    IGR J11203+4531 & 170.09 & 45.53 & 3273 \\
    IGR J11321-5311 & 173.03 & -53.18 & 545 \\
    1RXS J121324.5-601458 & 183.35 & -60.25 & 3271 \\
    IGR J12470-5407 & 191.75 & -54.13 & 3256 \\
    GB971227 & 194.37 & 59.27 & 3 \\
    1RXS J130737.8-425940 & 196.19 & -42.72 & M10  \\
    IGR J13186-6257 & 199.65 & -62.95 & 1539 \\
    CSS100413:132854+174318 & 202.23 & 17.72 & 2652 \\
    FERMI J1350-1140 & 207.51 & 11.68 & 3788 \\
    IGR J14003-6326 & 210.09 & -63.49 & 810 \\
    \hline
    IGR J14043-6148 & 211.20 & 61.81 & 3184 \\
    IGR J14298-6715 & 217.46 & -67.26 & 810 \\
    IGR J14466-3352 & 221.65 & 33.87 & 3065 \\
    IGR J14488-4008 & 222.21 & -40.14 & 3290 \\
    PMN J1508-4953 & 227.16 & -49.88 & 4167 \\
    TXS 1530-131 & 233.19 & -13.32 & 3579 \\
    IGR J15391-5307 & 234.77 & -53.12 & 3293 \\
    IGR J15529-5029 & 238.23 & -50.49 & 1539 \\
    IRAS F15596-7245 & 241.33 & -72.90 & 3271 \\
    IGR J16293-4603 & 247.28 & -46.06 & 1774 \\
    IRAS 16254-4557 & 247.28 & -46.06 &  \\
    IGR J16316-4028 & 247.90 & -40.47 & 253 \\
    XTE J1637-498 & 249.26 & -49.86 & 1704 \\
    IGR J16374-5043 & 249.34 & 50.73 & 2809 \\
    IGR J16413-4046  & 250.34 & -40.78 & 2731 \\
    IGR J16443+0131 & 251.08 & 1.52  & 3078 \\
    Unnamed AGILE Source & 252.90 & 53.70 & 3862 \\
    IGR J16558-4150 & 253.95 & -41.83 & 987 \\
    XTE J1704-445 & 256.13 & -44.53 & 1164 \\
    IGR J17062-6143 & 256.56 & -61.73 & 1840 \\
    FERMI J1717-5156 & 259.39 & -51.93 & 4023 \\
    IGR J17177-3656 & 259.41 & 36.94 & 3223 \\
    TXS 1715-354 & 259.74 & -35.55 & 2627 \\
    XTE J1719-291 & 259.82 & -29.07 & 1442 \\
    XTE J1726-476 & 261.71 & -47.64 & 623 \\
    SWIFT J1729.9-3437 & 262.54 & -34.61 & 2749 \\
    IGR J17354-3255 & 263.85 & -32.94 & 2019 \\
    IGR J17379-3747 & 264.48 & -37.78 & 1714 \\
    IGR J17419-2802 & 265.48 & -28.03 & 616 \\
    1LC 358.439-0.211 & 265.67 & -30.38 & 3273 \\
    IGR J17427-7319 & 265.68 & -73.33 & 3391 \\
    XTE J1743-363 & 265.75 & -36.35 & 332 \\
    IGR J17448-3232 & 266.16 & -32.54 & 1323 \\
    SWIFT J174535.5-285921 & 266.40 & 28.99 & 3472 \\
    IGR J17464-2811 & 266.82 & -28.18 & 970 \\
    IGR J17473-2721 & 266.83 & -27.34 & 1461 \\
    IGR J17487-3124  & 267.20 & -31.41 & 1332 \\
    \hline
    IGR J17494-3030 & 267.35 & -30.50 & 3984 \\
    IGR J17507-2856 & 267.69 & -28.95 & 342 \\
    IGR J17520-6018 & 268.01 & -60.31 & 2975 \\
    SAX J1805.5-2031 & 271.39 & -20.51 & 84 \\
    IGR J18175-1530 & 274.39 & -15.51 & 1248 \\
    IGR J18284-0345 & 277.13 & -3.76 & 1981 \\
    IGR J18325-0756 & 278.12 & -7.95 & 154 \\
    XTE J1837+037 & 279.16 & 3.68  & 3684 \\
    IGR J18371+2634 & 279.27 & 26.57 & 3185 \\
    IGR J18457+0244 & 281.42 & 2.75  & 3078 \\
    IGR J18462-0223 & 281.57 & -2.39 & 1319 \\
    IGR J18532+0416 & 283.29 & 4.27  & 3256 \\
    IGR J19094+0415 & 287.41 & 4.25  & 3361 \\
    IGR J19112+1358  & 287.80 & 13.98 & 2298 \\
    PMN J1913-3630 & 288.34 & -36.51 & 2966 \\
    IGR J19203+1328 & 290.15 & 13.47 & 3361 \\
    SWIFT J1922.7-1716 & 290.65 & -17.28 & 669 \\
    IGR J19284+0107 & 292.10 & 1.12  & 3125 \\
    GB980425 & 293.73 & -52.83 & 15 \\
    1RXS J200627.1+563309 & 301.61 & 56.55 & 3639 \\
    IGR J20188+3647 & 304.70 & 36.79 & 873 \\
    SWIFT J2037.2+4151 & 309.30 & 41.85 & 3272 \\
    SWIFT J2058.4+0516 & 314.58 & 5.23  & 3384 \\
    AGL 2103+5630 & 315.80 & 56.50 & 3544 \\
    IGR J21117+3427 & 317.95 & 34.46 & 873 \\
    GB980515 & 319.35 & -67.26 & 22 \\
    IGR J21441+4640 & 326.02 & 46.68 & 2975 \\
    NVSS J222329+010226 & 335.23 & 1.04  &  \\
    PMN J2250-2806 & 342.69 & -28.11 & 3787 \\
    PMN J2258-8246 & 343.23 & -83.04 & A10c  \\
    AGL 2302-3251 & 345.38 & -32.85 & 3357 \\
    IGR J23558-1047 & 358.94 & -10.79 & 3391 \\
    \hline

\end{longtable}
\end{center}
\vspace{-50pt}
{\small -- The complete sample of 102 sources used in this study. The `ATel \#' column indicates the Astronomer's Telegram from which the source was taken. Sources pulled from literature are referenced. The names and coordinates are as indicated in the original ATel or paper. The references are as follows: M10 -- Mahoney et al. 2010, M11 -- Miller et al. 2011, A10c -- Abdo et al 2011c. Sources without a specific reference were suggested by colleagues as being potentially interesting.}

\end{document}